\documentclass[11pt]{book}

\usepackage[dvips]{epsfig}
\usepackage{amssymb}
\usepackage{multirow}
\usepackage{amsbsy}
\usepackage{pstricks}

\catcode`\@=11

 \def\@normalsize{\@setsize\normalsize{13pt}\xipt\@xipt
   \abovedisplayskip 11pt plus3pt minus6pt
   \belowdisplayskip \abovedisplayskip
   \abovedisplayshortskip \z@ plus3pt
   \belowdisplayshortskip 6.6pt plus3.5pt minus3pt}

 \def\small{\@setsize\small{12pt}\xipt\@xipt
   \abovedisplayskip 10pt plus2pt minus5pt
   \belowdisplayskip \abovedisplayskip
   \abovedisplayshortskip \z@ plus3pt
   \belowdisplayshortskip 6pt plus3pt minus3pt
   \def\@listi{\topsep 6pt plus 2pt minus 2pt
     \parsep 3pt plus 2pt minus 1pt
     \itemsep \parsep}}

 \def\footnotesize{\@setsize\footnotesize{10pt}\ixpt\@ixpt
   \abovedisplayskip 8pt plus 2pt minus 4pt
   \belowdisplayskip \abovedisplayskip
   \abovedisplayshortskip \z@ plus 1pt
   \belowdisplayshortskip 4pt plus 2pt minus 2pt
   \def\@listi{\topsep 4pt plus 2pt minus 2pt
      \parsep 2pt plus 1pt minus 1pt
      \itemsep \parsep}}

 \def\scriptsize{\@setsize\scriptsize{9.5pt}\viiipt\@viiipt}
 \def\tiny{\@setsize\tiny{7pt}\vipt\@vipt}
 \def\large{\@setsize\large{14pt}\xiipt\@xiipt}
 \def\Large{\@setsize\Large{18pt}\xivpt\@xivpt}
 \def\LARGE{\@setsize\LARGE{22pt}\xviipt\@xviipt}
 \def\huge{\@setsize\huge{25pt}\xxpt\@xxpt}
 \def\Huge{\@setsize\Huge{30pt}\xxvpt\@xxvpt}

\def\section{\@startsection {section}{1}{\z@}%
{-1.5\baselineskip plus-1pt minus-3pt}{1\baselineskip plus1pt minus2pt}%
{\centering\normalsize\bf}}
\def\subsection{\@startsection{subsection}{2}{\z@}%
{-1\baselineskip plus-1pt minus-2pt}{1\baselineskip plus1pt minus2pt}%
{\normalsize\sc\noindent}}
\def\subsubsection{\@startsection{subsubsection}{3}{\z@}%
{-1\baselineskip plus-1pt minus-2pt}{1sp}{\normalsize\it\noindent}}
\def\paragraph{\@startsection{paragraph}{4}{\z@}%
{1\baselineskip plus1pt minus2pt}{-1em}{\normalsize\it\noindent}}
\let\subparagraph=\paragraph

\def\tableofcontents{\@restonecolfalse\if@twocolumn\@restonecoltrue
\onecolumn\fi\OSIDcont\@starttoc{con}\if@restonecol\twocolumn\fi}

\def\l@section{\@dottedtocline{1}{0em}{.66em}}

\def\thebibliography#1{\section*{{Bibliography}\@mkboth
 {BIBLIOGRAPHY}{BIBLIOGRAPHY}}\footnotesize\rm\list
 {[\arabic{enumi}]}{\settowidth\labelwidth{[#1]}\leftmargin\labelwidth
 \advance\leftmargin\labelsep\usecounter{enumi}}
 \def\newblock{\hskip .11em plus .33em minus -.07em}
 \sloppy\clubpenalty4000\widowpenalty4000
 \sfcode`\.=1000\relax}

\def\ps@myheadings{\let\@mkboth\@gobbletwo
\def\@oddhead{\hfil{\footnotesize\rm\rightmark}\hfil}
\def\@evenhead{\hfil{\footnotesize\rm\leftmark}\hfil}
\def\@oddfoot{\hfil{\footnotesize\sf\artid-\thepage}\hfil}
\def\@evenfoot{\hfil{\footnotesize\sf\artid-\thepage}\hfil}
\def\sectionmark##1{}\def\subsectionmark##1{}}

\def\@copyrighthead{}

\def\artid{0000001}
\def\Year{2008}        %
\newcounter{paPer}     %
\def\EndpagE{\expandafter\pageref{\the\value{paPer}OpSy}}

\def\ps@osiD{\let\@mkboth\@gobbletwo
\def\@oddhead{\@copyrighthead}
  \def\@oddfoot{\hfil{\footnotesize\sf\artid-\thepage}\hfil}
  \def\@evenhead{}\let\@evenfoot\@oddfoot}

\def\cite{\@ifnextchar [{\@tempswatrue\@Rcitex}{\@tempswafalse\@Rcitex[]}}

\def\@Rcitex[#1]#2{\if@filesw\immediate\write\@auxout{\string\citation{#2}}\fi
  \def\@citea{}\@cite{\@for\@citeb:=#2\do
    {\@citea\def\@citea{,\penalty\@m\,}\@ifundefined
       {b@\the\value{paPer}R\@citeb}{{\bf ?}\@warning
       {Citation `\@citeb' on page \thepage \space undefined}}%
\hbox{\csname b@\the\value{paPer}R\@citeb\endcsname}}}{#1}}

\long\def\@caption#1[#2]#3{\par\addcontentsline{\csname
  ext@#1\endcsname}{#1}{\protect\numberline{\csname
  the#1\endcsname}{\ignorespaces #2}}\begingroup
    \@parboxrestore
    \small                                        
    \@makecaption{\csname fnum@#1\endcsname}{\ignorespaces #3}\par
  \endgroup}

\newtoks\@stequation

\def\subequations{\refstepcounter{equation}%
\edef\@savedequation{\the\c@equation}%
\@stequation=\expandafter{\theequation}
\edef\@savedtheequation{\the\@stequation}
\edef\oldtheequation{\theequation}%
\setcounter{equation}{0}%
\def\theequation{\oldtheequation\alph{equation}}}%

\def\endsubequations{%
\setcounter{equation}{\@savedequation}%
\@stequation=\expandafter{\@savedtheequation}%
\edef\theequation{\the\@stequation}\global\@ignoretrue}

\catcode`\@=12

\let\Rlabel=\label
\let\Rbibitem=\bibitem
\let\Rref=\ref
\let\Rpageref=\pageref
\def\label#1{\expandafter\Rlabel{\the\value{paPer}R#1}}
\def\bibitem#1{\expandafter\Rbibitem{\the\value{paPer}R#1}}
\def\ref#1{\expandafter\Rref{\the\value{paPer}R#1}}
\def\pageref#1{\expandafter\Rpageref{\the\value{paPer}R#1}}

\def\thesection{\arabic{section}.}

\def\YYMm{\rule{0ex}{4em}}
\newtoks\TITsi
\newtoks\TITsii

\def\title#1{\def\TITs{\LARGE{\raggedright\noindent\YYMm #1%
\vskip8pt\par}}}

\def\author#1{\autMM{#1}\def\LHD{#1}}
\def\and{{\rm\lowercase{and}}}

\def\autMM#1{\TITsii={\vskip10pt\par\normalsize\rm\noindent #1\par}%
\TITsi=\expandafter{\TITs}\edef\TITs{\the\TITsi\the\TITsii}}

\def\address#1{\TITsii={\vskip6pt\par\footnotesize\sl\noindent #1\par}%
\TITsi=\expandafter{\TITs}%
\edef\TITs{\the\TITsi\the\TITsii}}

\def\received#1{\TITsii={\vskip10pt\par\small\rm\noindent(Received: #1)\par}%
\TITsi=\expandafter{\TITs}\edef\TITs{\the\TITsi\the\TITsii}}

\def\headtitle#1{\def\RHD{#1}}
\def\headauthor#1{\def\LHD{#1}}
\def\listas#1#2{\addcontentsline{con}{section}{{\sc #1: }{\rm #2}}}

\def\abst{{\bf Abstract.}}
\def\abstract#1{\TITs
       \vskip15pt\par\noindent
       {\footnotesize{\abst~} #1\vskip3pt\par}
       \markright{\RHD}
       \markboth{\LHD}{\RHD}}

\def\startpaper{%
       \cleardoublepage
       \setcounter{section}{0}
       \stepcounter{paPer}
       \setcounter{equation}{0}
       \setcounter{footnote}{0}
       \setcounter{figure}{0}
       \setcounter{table}{0}
       \def\theequation{\arabic{equation}}
       \def\thefootnote{\arabic{footnote}}
       \setcounter{defn}{0}
       \setcounter{thm}{0}
       \setcounter{lem}{0}
       \setcounter{prop}{0}
       \setcounter{rem}{0}
       \thispagestyle{osiD}}

\def\OSIDcont{\cleardoublepage\thispagestyle{empty}
       \markright{}\markboth{}{}
       \normalsize\rm
       \hspace*{\fill}{\large\rm
         Contents of the Volume \Volume, Number \Number}\hspace*{\fill}
       \par\vspace{1.5em}
       \par\noindent}

\def\endpaper{\expandafter\label{\the\value{paPer}OpSy}}

\def\emptyset{\mathchoice{\mbox{\normalsize\rm\O}}{\mbox{\normalsize\rm\O}}
{\mbox{\scriptsize\rm\O}}{\mbox{\tiny\rm\O}}}

\def\1{{\mathchoice{\rm 1\mskip-4mu l}{\rm 1\mskip-4mu l}%
{\rm 1\mskip-4.5mu l}{\rm 1\mskip-5mu l}}}

\def\varkappa{\mbox{\bBB\char 123}}

\def\longhookrightarrow{\lhook\joinrel\relbar\joinrel\rightarrow}

\def\longhookUp{\lower6pt\hbox{\rotatebox{90}{$\longhookrightarrow$}}}

\def\tr{\mathop{\rm tr}}

\newtheorem{thm}{\rm THEOREM}

\newtheorem{prop}{\rm PROPOSITION}

\newtheorem{defn}{\rm DEFINITION}

\def\theequation{\thesection\arabic{equation}}

\def\Myskip{\setlength{\baselineskip}{13pt}}

\def\text#1{\quad\mbox{\rm  #1 }\quad}

\def\DOInumber{}

\input xy
\xyoption{all}

\InputIfFileExists{psfig.sty}{\typeout{^^Jpsfig.sty inputed...ok}}{\typeout{^^JWarning: psfig.sty could not be found.^^J}}

\begin{document}

\def\artid{}
\def\Volume{}
\def\Number{}
\def\Year{}
\setcounter{page}{1}

\def\DOInumber{}

\startpaper

\newcommand{\pr}{\mathbb{P}}
\newcommand{\hh}{\mathfrak{h}}
\newcommand{\TT}{\mathcal{T}}
\newcommand{\LL}{\mathcal{L}}

\def\oper{{\mathchoice{\rm 1\mskip-4mu l}{\rm 1\mskip-4mu l}%
{\rm 1\mskip-4.5mu l}{\rm 1\mskip-5mu l}}}
\def\<{\langle}
\def\>{\rangle}
\def\theequation{\thesection\arabic{equation}}

\title{Quasi-stationary normal states for quantum Markov semigroups}
\author{Ameur Dhahri$^*$, Franco Fagnola$^*$, Federico Girotti$^*$ and Hyun Jae Yoo$^\dagger$}
\address{$^*$Department of Mathematics, Politecnico di Milano, Piazza Leonardo da Vinci 32, 20133 Milan, Italy}
\address{$^\dagger$ Department of Applied Mathematics and Institute for Integrated Mathematical Sciences, Hankyong National University, 327 Jungang-ro, Anseong-si, Gyeonggi-do 17579, Korea}
\headauthor{Ameur Dhahri, Franco Fagnola, Federico Girotti and Hyun Jae Yoo}
\headtitle{Quasi-stationary normal states for quantum Markov semigroups}
\received{}
\listas{Ameur Dhahri, Franco Fagnola, Federico Girotti and Hyun Jae Yoo}{Quasi-stationary normal states for quantum Markov semigroups}

\abstract{We introduce the notion of Quasi-Stationary State (QSS) in the context of quantum Markov semigroups that generalizes the one of quasi-stationary distribution in the case of classical Markov chains. We provide an operational interpretation of QSSs using the theory of direct and indirect quantum measurements. Moreover, we prove that there is a connection between QSSs and spectral properties of the quantum Markov semigroup. Finally, we discuss some examples which, despite their simplicity, already show interesting features.}

\Myskip



\section{Introduction}
\setcounter{equation}{0}

Let us consider a Markov chain $\mathbb{X}:=(X_t)_{t \geq 0}$ with state space indexed by a denumerable set $E$; given any probability density $\nu$ on $E$, we denote by $\pr_\nu$ the law of the process when $X_0$ is distributed according to $\nu$ and by $\mathbb{E}_{\nu}$ the corresponding expected value. In order to keep the notation light, for every state $x \in E$, we will use $\pr_x$ in place of $\pr_{\delta_x}$. We are interested in the situations when there exists a set of states $A \subset E$ which is absorbing, that means that the evolution hits $A$ in finite time almost surely and, once the process ends in $A$, it stays there forever. More precisely, if we define $T:=\inf\{t \geq 0: X_t \in A\}$, we assume the following:
\begin{enumerate}
\item for every $y \in A$ and $t \geq 0$, $\pr_y(X_t \in A)=1$ (\textbf{$A$ is closed}) and
\item for every $x \in E$, $\pr_x(T<+\infty)=1$ (\textbf{$A$ is absorbing}).
\end{enumerate}
A natural object of study is the behaviour of the Markov chain before hitting $A$ and a key concept in this regard is the one of Quasi-Stationary Distribution (QSD): a probability density $\nu$ supported on $A^C:=E\setminus A$ is said to be \textit{quasi-stationary} if it satisfies the following condition: for every $t \geq 0$ and for every $x \in A^C$ one has
\begin{equation} \label{eq:qsd}
\pr_\nu(X_t =x\,|\,T>t)=\nu(x).
\end{equation}
Condition in Eq. \ref{eq:qsd} states that $\nu$ is a stationary distribution when conditioning on the fact that the process does not hit $A$.

There is a wide literature regarding the study of Markov processes conditioned to `survive' and QSDs (see \cite{CMM} and references therein). A first fundamental property is that if $\nu$ is a QSD, then the survival probability $\pr_\nu(T>t)$ is equal to $e^{-\alpha t}$ for some positive constant $\alpha$; this provides a necessary condition for the existence of QSDs. A deeper result connects QSDs with spectral properties of the semigroup corresponding to $\mathbb{X}$, enabling to find simple criteria to ensure the existence and uniqueness of QSDs.

The purpose of this note is to consider the corresponding problem for Quantum Markov Semigroups (QMSs), which are among the most used mathematical models in the description of the evolution of open quantum systems. Aside from a purely mathematical interest, this work has two further motivations: closed and absorbing subspaces are central objects in the theoretical fundation of engineering and control of quantum system (see for instance \cite{TLCV12} and \cite{BPT17}). Moreover, the connection between QSS and the spectrum of the infinitesimal generator of the QMS provides a physical interpretation to some of the elements in the spectrum of the generator. The latter, in itself, has recently attracted some interest in the recent literature (see \cite{DLTCZ19,FL25,TCDZ23} and the references therein).

The main results of the paper characterize QSSs in terms of repeated measurements (Proposition \ref{prop:OP}), eigenvectors corresponding to a real eigenvalue of the infinitesimal generator (Theorem \ref{th:PF}). In addition, Theorem \ref{thm:exun} shows existence and uniqueness of QSS, for finite dimensional systems, as an application of the Perron-Frobenius theorem.

The structure of the paper is the following one: in Section \ref{sec:NP} we set the notation and introduce the basic concepts and results that are necessary in order to understand the content of this paper. Section \ref{sec:DOI} is devoted to provide a definition of Quasi-Stationary State (QSS) that generalizes the one of QSD and to suggest a possible operational interpretation. In Section \ref{sec:ST} we draw the connection between QSS and spectral properties of the QMS. Finally, in Section \ref{sec:EX} we present some examples of the theory treated in this work.

\section{Absorbing and subharmonic projections} \label{sec:NP}

We consider a quantum system described by a separable Hilbert space $\hh$ and we use $L^1(\hh)$ and $B(\hh)$  to denote trace class and bounded linear operators acting on $\hh$, respectively. A Quantum Markov Semigroup (QMS) is a collection of bounded linear operators $\TT_t:B(\hh) \rightarrow B(\hh)$ indexed by time $t \in [0,+\infty)$ such that
\begin{enumerate}
\item for every $t \geq 0$, $\TT_t$ is normal and completely positive,
\item for every $t \geq 0$, $\TT_t$ is unital, i.e. $\TT_t(\mathbf{1})=\mathbf{1}$,
\item for every $s,t \in [0,+\infty)$, $\TT_{t+s}=\TT_t\TT_s$,
\item $\TT_0={\rm Id}$,
\item $t \mapsto \TT_t$ is pointwise ${\rm w}^*$-continuous.
\end{enumerate}

For the sake of simplicity, we choose to stick to $B(\hh)$, however we point out that the theory developed in this paper holds for QMSs acting on general ${\rm W}^*$-algebras. $\TT:=(\TT_t)_{t \geq 0}$ describes the evolution of the system in the Heisenberg picture; the evolution of states corresponds to the collection of predual maps $\TT_{t*}$, which are uniquely determined by the follwing condition:
$${\rm tr}(x\TT_t(y))={\rm tr}(\TT_{t*}(x)y), \quad  x \in L^1(\hh), \, y \in B(\hh).$$
$\TT_*$ is a strongly continuous semigroup of completely positive and trace preserving maps.

The corresponding notion of closed set of states in the non-commutative framework is the support of a subharmonic projection (\cite{FP09}).

\begin{defn}[Subharmonic projection]
An orthogonal projection $p_0$ is said to be subharmonic for $\TT$ if $\TT_t(p_0) \geq p_0$ for every $t \geq0 $.
\end{defn}
One can easily see that $\TT_t(p_0^\perp) \leq p_0^\perp$ and $p_0^\perp$ is called superharmonic. This notion corresponds to the one of closed set of states for classical stochastic process because $p_0$ is subharmonic if and only if its range $\mathfrak{k}$ has the following property (Proposition II.1 in \cite{FR02}): if a state $\rho$ is supported in $\mathfrak{k}$, then $\TT_{t*}(\rho)$ is supported in $\mathfrak{k}$ as well for every $t \geq 0$.

Moreover, the following holds true: for every $x \in B(\hh)$ and for every $t \geq 0$, one has
$$ \TT_{t}(p_0^\perp x p_0^\perp)=p_0^\perp\TT_{t}(p_0^\perp x p_0^\perp)p_0^\perp.$$
Therefore, identifying $p_0^\perp B(\hh) p_0^\perp$ and $p_0^\perp L^1(\hh) p_0^\perp$ with $B(\mathfrak{k}^\perp)$ and $L^1(\mathfrak{k}^\perp)$, respectively, (where $\mathfrak{k}$ is the range of $p_0$), one can see that the collection of maps $\widehat{\TT}_t$ defined as
$$B(\mathfrak{k}^\perp) \ni x \mapsto \widehat{\TT}_t(x):=\TT_t(x) \in B(\mathfrak{k}^\perp)$$
is a pointwise ${\rm w}^*$-continuous semigroup of normal, completely positive, sub-unital (i.e. $\widehat{\TT}_t(p_0^\perp) \leq p_0^\perp$) linear maps. Its predual is given by
$$L^1(\mathfrak{k}^\perp) \ni x \mapsto \widehat{\TT}_{t*}(x):=p_0^\perp\TT_{t*}(x)p_0^\perp \in L^1(\mathfrak{k}^\perp). $$

The last notion we need is the quantum counterpart of an absorbing or attractive set of states.

\begin{defn}[Absorbing projection] \label{defn:abs}
We say that a subharmonic projection $p_0$ is absorbing if for every state $\rho$ one has that
$$\lim_{t \rightarrow +\infty} {\rm tr}(\TT_{t*}(\rho)p_0)=1.$$
\end{defn}

One can show (see \cite{CG}) that the ${\rm w}^*$-limit of the net $\TT_{t}(p_0)$ is well defined and it is called the absorption operator corresponding to $p_0$, which we will denote by $A(p_0)$. The condition in Definition \ref{defn:abs} can be rephrased as $A(p_0)=\mathbf{1}$. One can check that a projection is absorbing if and only if its range is globally asymptotically stable in the sense of \cite{TLCV12}.

\section{Definition and operational interpretation} \label{sec:DOI}
The starting point for defining a Quasi-Stationary State (QSS) is the formulation of Eq. \ref{eq:qsd} in terms of the Markov semigroup; we recall that the semigroup corresponding to a Markov chain $\mathbb{X}=(X_t)_{t \geq 0}$ with state space $E$ is defined as
$$\ell^\infty(E) \ni f(x) \mapsto T_t(f)(x):=\mathbb{E}_x[f(X_t)] \in \ell^\infty(E),
$$
where $\ell^\infty(E):=\{f:E \rightarrow E: \sup_{x \in E}|f(x)|<+\infty\}.$ The predual semigroup $\TT_{t*}$ acts on densities $\ell^1(E):=\{f:E \rightarrow E: \sum_{x \in E}|f(x)|<+\infty\}$ and is uniquely determined by the duality condition:
$$\sum_{x \in E} f(x)\TT_{t*}(g)(x)=\sum_{x \in E} \TT_t(f)(x)g(x), \quad f \in \ell^\infty(E), \, g \in \ell^1(E).$$
First of all, notice that
$$\pr_\nu(X_t=x|T>t)=\frac{\pr_\nu(X_t=x,T>t)}{\pr_\nu(T>t)}=\frac{\pr_\nu(X_t=x)}{\pr_\nu(T>t)}=\frac{T_{t*}(\nu)(x)}{\pr_\nu(T>t)},$$
since $x \in A^C$ and, therefore, due to the fact that $A$ is a closed set of states, one has $\{X_t=x\} \subseteq \{T>t\}$. One can express $\pr_\nu(T>t)$ as well in terms of the semigroup:
$$\pr_\nu(T>t)=\sum_{x \in A^C}\pr_\nu(X_t=x)=\sum_{x \in A^C} T_{t*}(\nu)(x)=T_{t*}(\nu)(\chi_{A^C}).$$
We use $\chi_{A^C}$ to denote the indicator function of $A^C$. Using that Eq. \ref{eq:qsd} holds for every $x \in A^C$ and putting together what we just observed, we can rephrase the definition of a QSD as
$$\frac{T_{t*}(\nu)\cdot \chi_{A^C}}{ T_{t*}(\nu)(\chi_{A^C})}=\nu, \quad \forall t \geq 0.$$
The last equation suggests the following generalization of the notion of QSD to the noncommutative setting.

\begin{defn}[Quasi-Stationary State]
A state $\nu$ is said to be a Quasi-Stationary State (QSS) for $\TT$ with respect to the subharmonic projection $p_0$ if the following holds true:
\begin{equation} \label{eq:defqss}
\frac{p_0^\perp\TT_{t*}(\nu)p_0^\perp}{{\rm tr}(\TT_{t*}(\nu)p_0^\perp)}=\nu, \quad \forall t \geq 0.
\end{equation}
\end{defn}

Even in the quantum case we can intepret the LHS of Eq. \ref{eq:defqss} as a conditional state: let us assume that the system is initially prepared in the state $\rho$, that it evolves according to $\TT_*$ for a certain time $t$ and then we measure the PVM given by $\{p_0, p_0^\perp\}$ which determines whether the system has reached the range of $p_0$ or not (we denote the outcomes by $0$ and $1$, repsectively). Born rule states that the result of the measurement is $0$ with probability ${\rm tr}(\TT_{t*}(\rho)p_0)$ and, in this case, the conditional state of the system is
$$\frac{p_0\TT_{t*}(\nu)p_0}{{\rm tr}(\TT_{t*}(\nu)p_0)};$$
on the other hand, the measurement outcome is $1$ with probability ${\rm tr}(\TT_{t*}(\rho)p_0^\perp)$ and, in this case, the conditional state of the system is
$$\frac{p_0^\perp\TT_{t*}(\nu)p_0^\perp}{{\rm tr}(\TT_{t*}(\nu)p_0^\perp)},$$
which is exactly the LHS of Eq. \ref{eq:defqss}.

We will now provide two equivalent definitions of QSSs in terms of repeated direct and continuous indirect measurements; before doing that, we need to introduce the evolution of the system when we are able to detect indirectly whether the system is in the range of $p_0^\perp$. We will refer to \cite{BG09,Da76,GvHGC22,SD81} for the physical derivation of the evolution of the system state under indirect measurements and we will include the rigorous definition of the mathematical objects involved in Appendix A (an introduction of quantum counting processes can also be found in \cite{GGG23,MK03} and references therein). The conditional state of the system is decribed by the stochastic process $\rho_t$: jumps caused by indirect measurements occur according to a counting process with random intensity given by $\tr(\rho_{t_-}p_0^\perp)$. If a jump occurs at a certain time $t$, the state undergoes the following transformation:
\[
\rho_{t_-} \mapsto \frac{p_0^\perp \rho_{t_-} p_0^\perp}{{\rm tr}\left(\rho_{t_-}p_0^\perp\right)}.
\]
In between jumps the state evolves deterministically according to the strongly continuous contraction semigroup $\mathcal{S}_{t*}$ generated by
\begin{equation}\label{eq:ssem}\LL_{0*}(\rho):=\LL_*(\rho)+{\cal J}_*(\rho), \quad {\cal J}_*(\rho):=-\frac{1}{2}\{p_0^\perp,\rho\},
\end{equation}
where $\{\cdot, \cdot\}$ is the anticommutator. ${\cal J}_*$ is bounded, however $\LL_*$ is possibly unbounded and so is $\LL_{0*}$; in any case $D(\LL_{0*})=D(\LL_*)$ and Eq. \ref{eq:ssem} makes sense for $\rho \in D(\LL_*)$. Let us denote by $T_1,\dots, T_n,\dots$ the random times at which jumps occur ($T_n$ is set to $+\infty$ if less then $n$ jumps are observed).
\begin{prop} \label{prop:OP}
Let $\nu$ be a state; the following statements are equivalent:

\begin{enumerate}
\item For all $t \geq 0$,
$$\frac{p_0^\perp\TT_{t*}(\nu)p_0^\perp}{{\rm tr}(\TT_{t*}(\nu)p_0^\perp)}=\nu,$$
\item For all $n \in \mathbb{N}$, for all $t_1, \dots, t_n \in [0,+\infty)$
$$\frac{p_0^\perp\TT_{t_n*}(p_0^\perp\TT_{t_{n-1}*}(\cdots p_0^\perp\TT_{t_1*}(\nu)p_0^\perp \cdots )p_0^\perp)p_0^\perp}{{\rm tr}(\TT_{t_n*}(p_0^\perp\TT_{t_{n-1}*}(\cdots p_0^\perp\TT_{t_1*}(\nu)p_0^\perp \cdots )p_0^\perp)p_0^\perp)}=\nu,$$
\item If the system is initially in the state $\nu$, then $T_1, \, T_2-T_1, \dots, T_{n+1}-T_n,\dots$ are independent and identically distributed exponential random variables with parameter $1+\alpha$; moreover, for every $n \geq 0$,
$$\rho_{T_n}=\nu.$$
\end{enumerate}
\end{prop}

The proof of Proposition \ref{prop:OP} can be found in Appendix A.

\section{Connection with spectral theory} \label{sec:ST}

In this section we will draw a connection between QSSs and spectral properties of $\TT$. We recall the following notations:
$$\widehat{\TT}_{t}(x):=\TT_{t}(p_0^\perp x p_0^\perp), \quad \widehat{\TT}_{t*}(y):=p_0^\perp\TT_{t*}(y) p_0^\perp, \quad x \in B(\hh), \, y \in L^1(\hh)$$
for $t \geq 0$. We will denote by $\widehat{\LL}$ and $\widehat{\LL}_*$ their generators, respectively.

\begin{thm} \label{thm:SP}
The following are equivalent
\begin{enumerate}
\item $\nu$ is a QSS for $\TT$ with respect to $p_0$;
\item There exists $\alpha\geq 0$ such that $\widehat{\TT}_{t*}(\nu)=e^{-\alpha t} \nu$;
\item $\nu \in D(\widehat{\LL}_*)$ and $\widehat{\LL}_*(\nu) =-\alpha \nu$ for some $\alpha\geq 0$.
\end{enumerate}
If $p_0$ is absorbing, i.e. $A(p_0)=\mathbf{1}$, then $\alpha>0$.
\end{thm}

The proof of Theorem \ref{thm:SP} can be found in Appendix B. We will now restrict to the case when \textit{$\hh$ is finite dimensional}. In this case, the semigroup is uniformly continuous and the infinitesimal generator is a bounded operator. Moreover, from basic spectral theory one has that the existence of $\nu$ satisfying item $3.$ in Theorem \ref{thm:SP} is equivalent to the existence of $x \in B(\hh)$, $x=x^*$ such that $x=p_0^\perp x p_0^\perp$ and
\begin{enumerate}
\item[$3.^\prime$] $\LL(x)=-\alpha x$ for some $\alpha\geq 0$.
\end{enumerate}
Moreover, the existence of such and $x$ ad some further properties are ensured by quantum Perron-Frobenius theory. Let us first introduce the notion of irreducible positive map (\cite{EHK}).
\begin{defn}
A positive map $\mathcal{A}:B(\hh) \rightarrow B(\hh)$ is said to be irreducible if there is not any non-trivial projection $p$ such that
$\mathcal{A}(p) \leq a p$ for some $a \geq 0$.

A semigroup of positive maps $(\mathcal{A}_t)_{t \geq 0}$ is said to be irreducible if for every $t \geq 0$, $\mathcal{A}_t$ is irreducible.
\end{defn}

One can show that if $(\mathcal{A}_t)_{t \geq 0}$ is a QMS acting on a finite dimensional Hilbert space, the definition above of irreducibility coincide with having a unique faithful invariant state (Theorem II.1 in \cite{FR02}). If $(\mathcal{A}_t)_{t \geq 0}$ is a uniformly continuous semigroup of completely positive maps, then its generator is of the form $\mathcal{B}(x)=g^*x+xg+\sum_{i \in I}a_i^* x a_i$ where $\sum_{i \in I}a_i^*a_i$ converges in the strong operator topology. An equivalent formulation of irreducibility can be given in terms of $g$ and jump operators $a_i$'s: indeed, $(\mathcal{A}_t)_{t \geq 0}$ is irreducible if there are not non-trivial invariant subspaces for $g$ and $a_i$'s (the proof follows the line of Theorem III.1 in \cite{FR02} without all the difficulties due to the unboundedness of the infinitesimal generator).

For readers' convenience we report below the results from Perron-Frobenius theory and, more in general, some well known results in the theory of positivity preserving maps acting on finite dimensional functional and matrix spaces ($C^*$-algebras) that are of interest for this work. We refer to \cite{EHK} for the general case of $C^*$-algebras. We remark that the following Theorem holds when $\hh$ is finite dimensional.

\begin{thm}[Perron-Frobenius Theory] \label{th:PF}
Let $r$ be the spectral radius of $\Psi$. The following statements hold true.
\begin{enumerate}
    \item $r \in {\rm Sp}(\Psi)$ and there exists $x \in B(\hh)$, $x \geq 0$ such that $\Psi(x)=rx$.
\end{enumerate}
If $\Psi$ is irreducible, then one has some further results.
\begin{enumerate}
\item $r$ is a geometrically simple eigenvalue;
\item $x$ is strictly positive and is the unique positive eigenvector.
\end{enumerate}
\end{thm}

The following proposition is a consequence of what we have discussed so far.

\begin{thm} \label{thm:exun}
Let $\hh$ be finite dimensional. For every subharmonic projection $p_0$, there exists a QSS $\nu$ for $\TT$ with respect to $p_0$. Moreover, if $\widehat{\TT}$ is irreducible, then $\nu$ is unique.
\end{thm}

We remark that the spectral radius of the restriction of the semigroup given by a superharmonic projection has already appeared in connection with large deviations of the position process in homogeneous open quantum random walks (see Proposition 5.5 in \cite{CGH22}) and in the nested phase decomposition in \cite{CT}.

\section{Two-qubit examples} \label{sec:EX}

Each quasi-stationary distribution of a finite state continuous time Markov chain can also be seen as a quasi-stationary
state for a quantum Markov semigroup. In this section we present an illustrative example in which the set of
QSS is genuinely non-commutative because it also contains densities that do not commute with each other.

Let $\mathfrak h=\mathbb C^2\otimes \mathbb C^2$ be the two-qubit Hilbert space.
Denote the canonical basis of $\mathbb C^2$ by $\{|0\rangle,|1\rangle\}$ with $|0\rangle =[1,0]^T$, $|1\rangle=[0,1]^T$.
Consider the operators
\[
\sigma_1^-=|0\rangle\langle 1|\otimes \mathbf{1},\quad \sigma_2^-
=\mathbf{1}\otimes |0\rangle \langle 1|,\quad \sigma_i^+=(\sigma_i^-)^*,\,\,i=1,2.
\]

We introduce the Hamiltonian $H:=\omega(\sigma_1^+\sigma_2^-+\sigma_1^-\sigma_2^+)/2$ with $\omega$ non-zero constant and we use the usual notation $[\cdot,\cdot]$,
$\{\cdot,\cdot\}$ for denoting the commutator and anti-commutator, respectively. In the sequel, we write the vector
$|i\rangle\otimes |j\rangle\in \mathfrak h$, $i,j\in \{0,1\}$, by $|ij\rangle$ for simplicity.
Denote $\mathfrak b=\{|00\rangle,|01\rangle,|10\rangle,|11\rangle\}$ a basis of $\mathfrak h$.

\subsection{Decay on one site only} 
Define a GKLS generator $\mathcal{L}$ on $\mathcal{B}(\mathfrak h)$ by
\begin{equation} \label{eq:GKSL}  
\mathcal L(x) = \mathrm{i}[H,x] \label{eq:GKSL}  
-\frac{1}{2}\left(\{\sigma_1^+\sigma_1^-, x\}-2 \sigma_1^+(x)\sigma_1^-\right).
\end{equation}

One immediately checks that $|00\rangle\langle 00|$ is an invariant density, so the projection
$p_0=|00\rangle\langle 00|$ is subharmonic (Theorem II.1 in \cite{FR02}) and we look for QSS with respect to $p_0$.

\begin{prop}\label{prop:two-qubit-QSS}
In the basis $\mathfrak{b}$,   QSSs  for the two-qubit system with respect to $p_0$ are the following
\begin{enumerate}
\item if $|\omega|\geq 1/2$,
\[
\nu=\left[\begin{array}{cccc}
0&0&0&0\\
0&1/2&x+\mathrm{i}(4\omega)^{-1}&0\\
0&x-\mathrm{i}(4\omega)^{-1}& 1/2 &0\\
0&0&0&0\end{array}\right], \quad x \in \mathbb{R}, \, |x|^2\leq (4\omega^2-1)/(16 \omega^2),
\]
corresponding to the eigenvalue $\alpha=1/2$;
\item if $|\omega| \leq 1/2$,
\[
\nu_+=\left[\begin{array}{cccc}
0&0&0&0\\
0&\alpha_-&\mathrm{i}\omega&0\\
0&-\mathrm{i}\omega& \alpha_+ &0\\
0&0&0&0\end{array}\right], \quad
\nu_-=\left[\begin{array}{cccc}
0&0&0&0\\
0&\alpha_+&\mathrm{i}\omega&0\\
0&-\mathrm{i}\omega& \alpha_- &0\\
0&0&0&0\end{array}\right],
\]
\sloppy corresponding to the eigenvalues $\alpha_+=\left (1+\sqrt{1-4\omega^2} \right )/2$ and $\alpha_-=\left (1-\sqrt{1-4\omega^2} \right )/2$, respectively.
\end{enumerate}
\end{prop}
$\nu_\pm$ are the only QSSs with extremal supports in the case $|\omega| \leq 1/2$, while for $|\omega|>1/2$, the elements with extremal supports are the QSSs corresponding to the choices $x=\pm \sqrt{(4 \omega^2-1)/(16 \omega^2)}$ (which we will call $\nu_\pm$ as well). Notice that $\nu_{\pm}$ are pure states, but they do not have orthogonal support in general: conversely to the set of stationary states of QMSs, for QSSs in general one cannot decompose the smallest subspace containing all the supports of QSSs into orthogonal supports of QSSs with minimal support.

\bigskip\noindent A possible interpretation of the above result is the following. If $\omega$ is big enough, then the rotation 
determined by the Hamiltonian term $\mathrm{i}[H,\cdot]$ of the GKLS generator is fast and the system has many possible QSSs but 
all with probability of observing it in $|01\rangle$ or $|10\rangle$ equal to $1/2$. As $\omega$ decreases and crosses the threshold $1/2$, 
the QSS are reduced to only two pure states with different probabilities of observing the system in $|01\rangle$ or $|10\rangle$. 
Clearly, the system undergoes a bifurcation as $\omega$ crosses the value $1/2$.

\subsection{Decay on both sites} Define a GKLS generator $\mathcal{L}$ on $\mathcal{B}(\mathfrak h)$ by
\begin{eqnarray}
\mathcal L(x_1\otimes x_2)& = &\mathrm{i}[H,x_1\otimes x_2] \label{eq:GKSL} \\
& - & \frac{1}{2}\sum_{l=1}^2\left(\{\sigma_l^+\sigma_l^-, x_1\otimes x_2\}-2 \sigma_l^+(x_1\otimes x_2)\sigma_l^-\right)
\nonumber
\end{eqnarray}The difference from the previous example is that in this model there is a jump operator acting on the second qubit as well.

$|00\rangle\langle 00|$ remains an invariant density for the modified model, so the projection
$p_0=|00\rangle\langle 00|$ is subharmonic and we look for QSS with respect to $p_0$.

\begin{prop}\label{prop:two-qubit-QSS2}
In the basis $\mathfrak{b}$, the only QSS $\nu$ for the two-qubit system with respect to $p_0$ is given by
\[
\nu = \frac{1}{2}\left( |01\rangle\langle 01|+|10\rangle\langle 10|\right)
\]
with eigenvalue $\alpha=1$.
\end{prop}

The proof of Propositions \ref{prop:two-qubit-QSS} and \ref{prop:two-qubit-QSS2} can be found in Appendix C.

\section{Conclusion and Outlook}

In this paper we defined a generalization of the notion of quasi-stationary states for quantum Markov semigroups, we provided an operational interpretation and we proved some first properties connecting the notion of QSS with the spectrum of the semigroup and about the existence and uniqueness of a QSS for finite dimensional QMSs. Natural lines for further investigations are, for instance, understanding the situation in which the model admits more than one QSSs and what class of eigenvalues (in the case of finite dimensional systems) of the infinitesimal generator correspond to QSSs.

QSSs can also be interpreted as states where a system can remain for some, possibly long, time, before moving to a different, stable state. In this sense, they can be thought of as metastable states. The relationship with metastability could also be investigated.

\section*{Acknowledgments}
AD, FF and FG are members of GNAMPA-INdAM.
AD, FF and FG  acknowledge the support of the MUR grant ``Dipartimento di Eccellenza 2023--2027'' of Dipartimento di Matematica, Politecnico di Milano and ``Centro Nazionale di ricerca in HPC, Big Data and Quantum Computing''.
HJY has been supported by the Korean government grant MSIT (No. RS-2023-00244129).

\section*{Appendix A} \label{app:A}
\def\theequation{A.\arabic{equation}}
\setcounter{equation}{0}

In this section, we will provide the proof of Proposition \ref{prop:OP} and a rigorous definition for the objects involved in the evolution of the system state conditional to the output of indirect measurements.

\textbf{Definition of the conditional state process.} First of all, let us consider $(\LL_{0*}, D(\LL_{0*}))$ as defined in Eq. \ref{eq:ssem} and
$$ \widetilde{\LL}_*(\cdot)=\LL_*(\cdot)+D[p_0^\perp](\cdot)=\LL_{0*}(\cdot)+p_0^\perp \cdot p_0^\perp, \quad D(\widetilde{\LL}_*)=D(\LL_*),$$
where $D[p_0^\perp](\rho)=-\frac{1}{2}(p_0^\perp \rho-2p_0^\perp \rho p_0^\perp + \rho p_0^\perp)$.

Notice that, using the theory of bounded perturbation (see Theorem 1.3 in Chapter III of \cite{EN}) and Lumer-Phillips Theorem (Theorem 3.15, Chapter II in \cite{EN})), one can easily see that $(\LL_{0*}, D(\LL_{0*}))$ and $(\widetilde{\LL}_*, D(\widetilde{\LL}_*))$ both generate a strongly continuous semigroup of contractions, which we will denote by $\mathcal{ S}_*$ and $\widetilde{\mathcal{S}}_*$, respectively.

Moreover, using the product formula in Theorem 3.30 in \cite{Da}, one can see that the semigroups are also completely positive and that the one corresponding to $\widetilde{\LL}_*$ is also trace preserving.

We will now define the stochastic process $\rho_t$. For $S>0$, let us define
$$\Omega_S=\bigcup_{n \geq 0} \Omega_S^n,$$
where $\Omega_0=\{\emptyset\}$ and
$$\Omega_S^n:=\{(t_1,\dots, t_n) \in [0,S]^n:t_1\leq t_2 \cdots \leq t_n \}.$$
We can endow $\Omega_S^n$ with the $\sigma$-field inherited from $[0,S]^n$ (considered with the Lebesgue $\sigma$-field) and consider the disjoint union sigma-field on $\Omega_S$. We will denote by $\mu$ the measure which is equal to $1$ on $\Omega_S^0$ and coincide with the restriction of the Lebesgue measure on $\Omega_S^N$ for $n \geq 1$. Let us define the following random variables: for every $t \leq S$
$$
N_t(\emptyset)=0, \quad N_t((t_1,\dots, t_n))=\sum_{i=1}^{n}\chi_{[0,t]}(t_i)$$
and
$$\rho_t(\emptyset)=\mathcal{S}_t(\rho), \quad \rho_t((t_1,\dots, t_n))=\mathcal{S}_{t-t_{N_t}}(p_0^\perp\cdots {\cal S}_{t_2-t_1}(p_0^\perp \mathcal{S}_{t_1}(\rho)p_0^\perp)\cdots p_0^\perp).
$$
The last ingredient is the following probability measure: given any initial state $\rho$, we define
$$\frac{d\pr_{\rho,S}}{d\mu}(\omega)={\rm tr}(\rho_S(\omega)), \quad \omega \in \Omega_S.$$
From the properties of $\mathcal{S}$ and its explicit expression, one can easily check that $\frac{d\pr_{\rho,S}}{d\mu}\geq 0$. Notice that $\widetilde{\LL}_*(\cdot)=\LL_0*(\cdot)+p_0^\perp \cdot p_0^\perp$, therefore Theorem 1.10 in \cite{EN} allows one to write
\begin{eqnarray*}&&\pr_{\rho,S}(\Omega_S)={\rm tr}(\mathcal{S}_{S*}(\rho))\\
&&+\sum_{k \geq 1} \int_{0 \leq t_1 \leq \cdots \leq t_k \leq S} {\rm tr}(\mathcal{S}_{(t-t_k)*}(p_0^\perp\cdots p_0^\perp \mathcal{S}_{t_1 *}(\rho)p_0^\perp\cdots p_0^\perp))dt_1 \cdots dt_k\\
&&={\rm tr}(\widetilde{\mathcal{S}}_{S*}(\rho))=1.
\end{eqnarray*}

Moroever, with the same series expansion and using that $\widetilde{\mathcal{S}}$ is trace preserving, one can see that the family of probability measures corresponding to different values of $S$ are compatible and, hence, we can use Kolmogorov extension theorem to remove the finite time horizon.

\bigskip \textbf{Proof of Proposition \ref{prop:OP}.} $2. \Rightarrow 1.$ is trivial, while the reverse implication follows easily from the fact that for every state $\rho$, one has
$$p_0^\perp \TT_{t*}(p_0^\perp \rho p_0^\perp)p_0^\perp=p_0^\perp \TT_{t*}( \rho )p_0^\perp.$$

\bigskip $2. \Rightarrow 3.$ Notice that for every $t \geq 0$, one has
\begin{equation} \label{eq:pr1}\frac{p_0^\perp \mathcal{S}_{t*}(\nu)p_0^\perp}{{\rm tr}(\mathcal{S}_{t*}(\nu)p_0^\perp)}=\nu.
\end{equation}
Indeed, using Theorem 1.10 in \cite{EN} one has that
$$\mathcal{S}_{t*}=\TT_{t*}+\sum_{k \geq 1} \int_{0 \leq t_1 \leq \cdots \leq t_k \leq t} \TT_{(t-t_k)*}\mathcal{J}_*\cdots\mathcal{J}_* \TT_{t_1 *}dt_1 \cdots dt_k.$$
Therefore, since $p_0^\perp \TT_{t*}(p_0^\perp \rho p_0^\perp)p_0^\perp=p_0^\perp (\TT_{t*} \rho )p_0^\perp$ and $p_0^\perp {\cal J}_*(\rho)p_0^\perp=-p_0^\perp \rho p_0^\perp$, one has that
\begin{eqnarray}\label{eq:Sexp}
&&p_0^\perp \mathcal{S}_{t*}(\nu)p_0^\perp=p_0^\perp\TT_{t*}(\nu)p_0^\perp\\
&&+\sum_{k \geq 1} (-1)^k \int_{0 \leq t_1 \leq \cdots \leq t_k \leq t}p_0^\perp \TT_{(t-t_k)*}(p_0^\perp\cdots p_0^\perp \TT_{t_1 *}(\nu)p_0^\perp \cdots p_0^\perp)dt_1 \cdots dt_k.\nonumber\end{eqnarray}
and Eq. \ref{eq:pr1} follows from 2.. Therefore, by induction and using that
$$\rho_{T_{n+1}}=\frac{p_0^\perp \mathcal{S}_{(T_{n+1}-T_n)*}(\rho_{T_n})p_0^\perp}{{\rm tr}(\mathcal{S}_{(T_{n+1}-T_n)*}(\rho_{T_n})p_0^\perp)},$$
one has that
$$\rho_{T_n} \cdot \chi_{T_n <\infty} =\nu \cdot \chi_{T_n<+\infty},$$
where $T_0=0$.

Therefore, $\{T_{n+1}-T_n\}_{n \geq 0}$ are all independent and identically distributed random variables with survival function given by:
\begin{eqnarray*}
&&{\rm tr}(\mathcal{S}_{t*}(\nu)p_0^\perp)={\rm tr}(\TT_{t*}(\nu)p_0^\perp)\\
&&+\sum_{k \geq 1} (-1)^k \int_{0 \leq t_1 \leq \cdots \leq t_k \leq t}{\rm tr}(\TT_{(t-t_k)*}(p_0^\perp\cdots p_0^\perp \TT_{t_1 *}(\nu)p_0^\perp \cdots p_0^\perp))dt_1 \cdots dt_k\\
&&=e^{-\alpha t} \left (1 +\sum_{k \geq 1} \frac{(-t)^k}{k!} \right )=e^{-(1+\alpha)t},
\end{eqnarray*}
where we used Eq. \ref{eq:Sexp}. We recall that $e^{-\alpha t}={\rm tr}(\TT_{t*}(\nu)p_0^\perp).$ In particular, we get that $T_n$ is almost surely finite for every $n \geq 0$.

\bigskip $3. \Rightarrow 1.$ It follows easily from the fact that the range of $T_1$ is equal to $[0,+\infty]$.

\section*{Appendix B} \label{app:B}
\def\theequation{B.\arabic{equation}}
\setcounter{equation}{0}

\textbf{Proof of Theorem \ref{thm:SP}.}  One can immediately see that $2.$ and $3.$ are equivalent.

Moreover, $1.$ can be easily seen to be a direct implication of $2.$. The only non-trivial implication is $1. \Rightarrow 2.$: let us introduce the notation $f(t):={\rm tr}(\TT_{t*}(\nu)p_0^\perp)$; the continuity property of the semigroup implies that $f$ is continuous. Moreover, using the semigroup property together with the definition of QSS, one has that for every $t,s \geq 0$
\begin{eqnarray*}
{\rm tr}(\TT_{(t+s)*}(\nu)p_0^\perp)\nu&=&\TT_{(t+s)*}(\nu)=\TT_{t*}(\TT_{s*}(\nu))={\rm tr}(\TT_{s*}(\nu)p_0^\perp)\TT_{t*}(\nu)\\
&=&{\rm tr}(\TT_{s*}(\nu)p_0^\perp){\rm tr}(\TT_{t*}(\nu)p_0^\perp)\nu.
\end{eqnarray*}
Therefore, $f(t+s)=f(t)f(s)$ for every $t,s \geq 0$ and, calling $\alpha:=-\log(f(1))$, by standard arguments one can deduce that $f(t)=e^{-\alpha t}$ for rational $t$'s and extend to all arguments by continuity.

\bigskip If $p_0$ is absorbing, it means that
$$0={\rm tr}(\nu(\mathbf{1}-A(p_0)))=\lim_{t \rightarrow +\infty}{\rm tr}(\TT_{t*}(\nu)p_0^\perp)=\lim_{t \rightarrow +\infty}e^{-\alpha t},$$
therefore $\alpha>0$.

\section*{Appendix C} \label{app:C}
\def\theequation{C.\arabic{equation}}
\setcounter{equation}{0}

\noindent \textbf{Proof of Proposition \ref{prop:two-qubit-QSS}.}
A state $\nu$ supported on $p_0^\perp$ has a form:
\begin{equation}\label{eq:general_form}
\nu=\left[\begin{array}{cccc}0&0&0&0\\0&\nu_{11}&\nu_{12}&\nu_{13}\\0&\overline {\nu_{12}}&\nu_{22}&\nu_{23}\\0&\overline{\nu_{13}}&\overline{\nu_{23}}&\nu_{33}\end{array}\right].
\end{equation}
A direct computation shows that $\mathcal{L}_*(\nu)$ is equal to
\[
\left[ \begin{array}{cccc}
\nu_{22} & \nu_{23} & 0 & 0 \\
\overline{\nu_{23}} & \nu_{33}-\omega{\rm Im}\nu_{12}
 & -\frac{\nu_{12}}{2} +\frac{\mathrm{i}\omega(\nu_{11}-\nu_{22})}{2} & -\frac{\nu_{13}+\mathrm{i}\omega\nu_{23}}{2} \\
 0 & -\frac{\overline{\nu_{12}}}{2} - \frac{\mathrm{i}\omega(\nu_{11}-\nu_{22})}{2}
  & -\nu_{22} + \omega{\rm Im} \nu_{12} & -\nu_{23}-\frac{\mathrm{i}\omega\nu_{13}}{2} \\
  0 & -\frac{\overline{\nu_{13}}-\mathrm{i}\omega\overline{\nu_{23}}}{2}
  & -\overline{\nu_{23}} + \frac{\mathrm{i}\omega\overline{\nu_{13}}}{2} & -\nu_{33}
\end{array} \right]
\]
By Theorem \ref{thm:SP}, one finds QSS solving the system
\begin{eqnarray*}
  \nu_{33} -\omega\,{\rm Im{\nu_{12}}} &=& -\alpha \nu_{11} \\
  -\nu_{12} +\mathrm{i}\omega(\nu_{11}-\nu_{22}) &=& -2\alpha \nu_{12} \\
  -\nu_{13}-\mathrm{i}\,\omega\nu_{23} &=& - 2\alpha \nu_{13} \\
  -\nu_{22}+\omega\,{\rm Im}\nu_{12} &=& -\alpha \nu_{22} \\
  -2\nu_{23} -\mathrm{i}\,\omega \nu_{13} &=& - 2\alpha\nu_{23} \\
  -\nu_{33} &=& -\alpha\nu_{33}
\end{eqnarray*}
A quick look at the above equations shows that, for $\alpha=1$, we get (fourth equation) ${\rm Im}\nu_{12}=0$ whence $\nu_{33}=-\nu_{11}$ (first equation) and so, by positivity,
$\nu_{33}=\nu_{11}=0$ and $\nu_{12}=\nu_{13}=\nu_{23}=0$. However, the second equation implies that $0=\nu_{11}=\nu_{22}$ and we get that no state can be a QSS with rate $\alpha=1$.

In the other cases ($\alpha \neq 1$) $\nu_{33}=0$, by positivity $\nu_{13}=\nu_{23}=0$ and the above system reduces to
\begin{eqnarray*}
  -\omega\,{\rm Im{\nu_{12}}} &=& -\alpha \nu_{11} \\
  -\nu_{12} +\mathrm{i}\omega(\nu_{11}-\nu_{22}) &=& -2\alpha \nu_{12} \\
  -\nu_{22}+\omega\,{\rm Im}\nu_{12} &=& -\alpha \nu_{22} \\
\end{eqnarray*}
If we set $\alpha=1/2$, the system becomes
\begin{eqnarray*}
  -\omega\,{\rm Im{\nu_{12}}} &=& - \nu_{11}/2 \\
  \omega(\nu_{11}-\nu_{22}) &=& 0 \\
  -\nu_{22}+\omega\,{\rm Im}\nu_{12} &=& - \nu_{22}/2 \\
\end{eqnarray*}
The second equation and the fact that $\nu$ is a state imply that $\nu_{11}=\nu_{22}=1/2$ (where the last equality is due to normalization) and the first and third equations become equal to
$$ \omega {\rm Im}\,\nu_{12} = 1/4.$$
Therefore, we find the family of states
\[
\nu=\left[\begin{array}{cccc}
0&0&0&0\\
0&1/2&x+\mathrm{i}(4\omega)^{-1}&0\\
0&x-\mathrm{i}(4\omega)^{-1}& 1/2 &0\\
0&0&0&0\end{array}\right], \quad |x|^2\leq (4\omega^2-1)/(16 \omega^2),
\]
which is non-empty if $|\omega|\geq 1/2$.

Let us now consider the cases when $\alpha \neq 1,1/2$. In this case the second equation implies that $\nu_{12}=:ix$ is purely imaginary and we get
\begin{eqnarray*}
  \omega x &=& \alpha \nu_{11} \\
  \omega(\nu_{11}-\nu_{22}) &=& (1-2\alpha) x \\
  \omega x &=& (1-\alpha) \nu_{22}. \\
\end{eqnarray*}
Using the first and last equations, together with the fact that $\nu$ is a state, we obtain that
$$\nu_{11}=(1-\alpha), \quad \nu_{22}=\alpha, \quad \alpha >1.$$
Substituting in the first equation and second equations, we obtain
\begin{eqnarray*}
   x &=& \alpha (1-\alpha) \omega^{-1} \\
  x&=&\omega. \\
\end{eqnarray*}
The solutions for $\alpha$ are real if and only if $|\omega|\leq 1/2$ and, in this case, are exactly given by $\alpha_{\pm}$.

\bigskip \noindent\textbf{Proof of Proposition \ref{prop:two-qubit-QSS2}.}
A state $\nu$ supported on $p_0^\perp$ has a form:
\begin{equation}\label{eq:general_form}
\nu=\left[\begin{array}{cccc}0&0&0&0\\0&\nu_{11}&\nu_{12}&\nu_{13}\\0&\overline {\nu_{12}}&\nu_{22}&\nu_{23}\\0&\overline{\nu_{13}}&\overline{\nu_{23}}&\nu_{33}\end{array}\right].
\end{equation}
A direct computation shows that $\mathcal{L}_{*}(\nu)$ is equal to
\[
\left[
\begin{array}{cccc}
 \nu_{11}+\nu_{22} & \nu_{23} & \nu_{13} & 0 \\
 \overline{\nu_{23}} & (\nu_{33}-\nu_{11}) -\omega\,{\rm Im}(\nu_{12})
            & -\nu_{12}+\mathrm{i}\,\frac{\omega\,(\nu_{11}-\nu_{22})}{2} & -\frac{3\nu_{13}}{2}-\mathrm{i}\,\frac{\omega\,\nu_{23}}{2} \\
\overline{\nu_{13}} & -\overline{\nu_{12}}-\mathrm{i}\,\frac{\omega\,(\nu_{11}-\nu_{22}}{2})
& (\nu_{33}-\nu_{22})+\omega\,{\rm Im}(\nu_{12}) & -\frac{3\nu_{23}}{2}-\mathrm{i}\,\frac{\omega\,\nu_{13}}{2} \\
 0 & -\frac{3\overline{\nu_{13}}}{2}+\mathrm{i}\,\frac{\omega\,\overline{\nu_{23}}}{2}
 & -\frac{3\overline{\nu_{23}}}{2}+\mathrm{i}\,\frac{\omega\,\overline{\nu_{13}}}{2} & -2\nu_{33} \\
\end{array}
\right]
\]
By Theorem \ref{thm:SP}, one finds QSS solving the system
\begin{eqnarray*}
  (\nu_{33}-\nu_{11}) -\omega\,{\rm Im}(\nu_{12}) &=&  -\alpha \nu_{11} \\
  -\nu_{12}+\mathrm{i}\,\omega\,(\nu_{11}-\nu_{22})/2  &=&  -\alpha\nu_{12} \\
  -3\nu_{13}-\mathrm{i}\,\omega\,\nu_{23} & = & -2\alpha \nu_{13} \\
  (\nu_{33}-\nu_{22})+\omega\,{\rm Im}(\nu_{12}) & = & -\alpha \nu_{22} \\
  -3\nu_{23}-\mathrm{i}\,\omega\,\nu_{13} & = & -2\alpha \nu_{23} \\
  -2\nu_{33} & = & -\alpha \nu_{33}
\end{eqnarray*}
If $\alpha=2$, then the first and fourth equation become
\[
(\nu_{33}+\nu_{11}) -\omega\,{\rm Im}(\nu_{12}) = 0 \qquad  (\nu_{33}+\nu_{22})+\omega\,{\rm Im}(\nu_{12}) = 0
\]
and their sum leads to $2\nu_{33}+\nu_{11}+\nu_{22}=0$ which is not an acceptable solution for a QSS.
Therefore, we assume $\alpha\not=2$ which implies $\nu_{33}=0$ and, in turn, by positivity $\nu_{13}=\nu_{23}=0$.
The system becomes
\begin{eqnarray*}
  -\nu_{11} -\omega\,{\rm Im}(\nu_{12}) &=&  -\alpha \nu_{11} \\
  -2\nu_{12}+\mathrm{i}\,\omega\,(\nu_{11}-\nu_{22})  &=&  -2\alpha\nu_{12} \\
  -\nu_{22}+\omega\,{\rm Im}(\nu_{12}) & = & -\alpha \nu_{22}
\end{eqnarray*}
Now the sum of the first and third equations yields $\alpha=1$ and the system can be rewritten as
\begin{eqnarray*}
  -\nu_{11} -\omega\,{\rm Im}(\nu_{12}) &=&  -\nu_{11} \\
  i\omega\,(\nu_{11}-\nu_{22})  &=&  0 \\
  -\nu_{22}+\omega\,{\rm Im}(\nu_{12}) & = & -\nu_{22}
\end{eqnarray*}
whence ${\rm Im}\nu_{12} =0$ and $\nu_{11}=\nu_{22}=1/2$. It turns out that
\[
\nu = \frac{1}{2}\left( |01\rangle\langle 01|+|10\rangle\langle 10|\right).
\]

\endpaper




\end{document}